\documentstyle[a4,12pt]{article}
\begin{document}
\parskip=5pt plus 1pt minus 1pt

\begin{flushright}
{\bf Preprint LMU-98-08}\\
{\bf hep-ph/9808272} \\
{July 1998}
\end{flushright}

\vspace{0.2cm}

\begin{center}
{\large\bf Large Leptonic Flavor Mixing and the Mass
Spectrum of Leptons} 
\end{center}

\vspace{0.4cm}
\begin{center}
{\bf Harald Fritzsch} 
\footnote{E-mail: bm@hep.physik.uni-muenchen.de}
~ {and} ~
{\bf Zhi-zhong Xing}
\footnote{E-mail: xing@hep.physik.uni-muenchen.de} \\
{\it Sektion Physik, Universit\"at M\"unchen,
Theresienstrasse 37, 80333 M\"unchen, Germany} 
\end{center}

\vspace{1.9cm}
\begin{abstract}
Implications of a simple model for the mass generation
of leptons are studied, in particular for the upcoming
long-baseline neutrino experiments. 
The flavor mixing angles are large (nearly maximal).
The probability for the long-baseline
$\nu_\mu \leftrightarrow \nu_e$ oscillation is 
predicted to be about $1\%$.
\end{abstract}

\newpage

Recently the Super-Kamiokande Collaboration has reported new and
stronger evidence for the existence of the atmospheric neutrino
anomaly. The data particularly
favor an interpretation of the
observed muon-neutrino deficit by a $\nu_\mu \leftrightarrow 
\nu_\tau$ oscillation with the mass-squared difference 
$\Delta m^2_{\rm atm} \approx (0.5 \dots 6) \times 10^{-3} ~ {\rm eV}^2$
and the mixing factor $\sin^2 2 \theta_{\rm atm} > 0.82$ 
at the $90\%$ confidence level \cite{SK1}. The long-standing solar neutrino
deficit has also been confirmed in the Super-Kamiokande
experiment. Analyses of the energy shape and
day-night spectra of solar neutrinos favor the mechanism
of a long-wavelength vacuum oscillation with $\Delta m^2_{\rm sun} 
\approx 10^{-10} ~ {\rm eV}^2$ and $\sin^2 2\theta_{\rm sun}
\approx 1$ \cite{SK2}.
If these large mixing angles
$\theta_{\rm atm}$ and $\theta_{\rm sun}$ are finally confirmed,
one would have an indication that the physics
responsible for neutrino masses and leptonic flavor mixing might be
qualitatively different from that for the quark sector.

In 1996 a pattern of lepton mass matrices, based on
the approximate flavor democracy for charged leptons and the 
near mass degeneracy for neutrinos, was proposed by the 
present authors \cite{FX96}.
In this approach the flavor mixing matrix in the symmetry limit is
identical to the mixing matrix that one obtains in QCD for
the light pseudoscalar mesons in the limit of chiral 
symmetry. Note that the mixing between the mass eigenstates
$|\pi^0 \rangle$, $|\eta \rangle$, $|\eta' \rangle$ and 
the QCD eigenstates $|\bar{u} u\rangle$, $|\bar{d} d\rangle$,
$|\bar{s} s\rangle$ is caused by the gluon anomaly,
which at the same time leads to a strong mass hierarchy:
$M^2_{\pi^0}, M^2_{\eta} \ll M^2_{\eta'}$. Analogously the
mixing of lepton flavors arises from a strong mismatch
between the charged lepton and neutrino mass matrices. 
To leading order we find a constant flavor mixing matrix,
independent of the lepton masses,
linking the neutrino mass eigenstates 
to the neutrino flavor eigenstates:
\begin{equation}
\left ( \matrix{
\nu_e \cr
\nu_\mu \cr
\nu_\tau } \right ) \; =\; U_0 \left ( \matrix{
\nu_1 \cr
\nu_2 \cr
\nu_3 \cr } \right ) \; \; ,
\end{equation}
where
\begin{equation}
U_0 \; =\; \left ( \matrix{
\frac{1}{\sqrt{2}}	& -\frac{1}{\sqrt{2}}	& 0 \cr
\frac{1}{\sqrt{6}}	& \frac{1}{\sqrt{6}}	& -\frac{2}{\sqrt{6}} \cr
\frac{1}{\sqrt{3}}	& \frac{1}{\sqrt{3}}	& \frac{1}{\sqrt{3}} \cr
} \right ) \; .
\end{equation}
Therefore one obtains
$\sin^2 2\theta_{\rm atm} = 8/9$ and $\sin^2 2\theta_{\rm sun}
=1$. Both angles are in good agreement with current data on atmospheric
and solar neutrinos.
As discussed in Ref. \cite{FX96}, this constant 
algebraic mixing matrix
will receive small corrections, once the muon and electron
masses are taken into account. We emphasize that in our approach
the leptonic flavor mixing angles do not depend on the neutrino masses.

The present note aims at displaying the symmetry-breaking 
patterns of the lepton mass matrices more clearly and 
discussing their next-to-leading-order consequences
on flavor mixing angles as well as specific predictions for the upcoming
neutrino experiments. In particular it will be shown
that a small correction
to $U_0$, which makes its (1,3) element deviate slightly
from zero, may lead to observable effects in the future long-baseline
neutrino experiments. 

Let us start with the symmetry limits of 
the charged lepton and
neutrino mass matrices. In the flavor space in which 
charged leptons have the exact flavor democracy and 
neutrino masses are fully degenerate, the mass matrices
can be written as 
\footnote{We assume $CP$ invariance throughout this work.
The $CP$ parities of two of the three (Majorana-type)
neutrinos may be different and have a certain consequence
on the neutrinoless $\beta\beta$-decay. In this case
$M_{0\nu}$ takes the more general form
$M_{0\nu} = c_\nu {\rm Diag} \{ \eta_1, \eta_2, \eta_3 \}$
with $\eta_i = \pm 1$ and $|m_i| =m_0$ \cite{Smirnov}.}
\begin{eqnarray}
M_{0l} & = & \frac{c^{~}_l}{3} \left ( \matrix{
1	& 1	& 1 \cr
1	& 1	& 1 \cr
1	& 1	& 1 \cr } \right ) \; , \nonumber \\
M_{0\nu} & = & c_\nu \left ( \matrix{
1	& 0	& 0 \cr
0	& 1	& 0 \cr
0	& 0	& 1 \cr } \right ) \; , 
\end{eqnarray}
where $c^{~}_l = m_\tau$ and $c_\nu = m_0$ measure the mass scales
of charged leptons and neutrinos, respectively.
Note that the special case $c_\nu =m_0 = 0$ is not excluded. The 
matrices given in Eq. (3) exhibit the underlying 
$S(3)_{\rm L} \times S(3)_{\rm R}$ symmetry for the charged
leptons and the $S(3)$ symmetry for the neutrinos 
(see also Ref. \cite{FTY98}). We leave it
open at the moment whether the neutrino masses are
of Fermi-Dirac or Majorana type.
Of course there is no flavor mixing in these symmetry limits,
in which $m_e = m_\mu =0$ and $m_1 = m_2 = m_3 =m_0$ hold.
A simple {\it diagonal} breaking of the flavor democracy
for $M_{0l}$ and the mass degeneracy for $M_{0\nu}$,
as introduced in Ref. \cite{FX96}, may lead to phenomenologically
instructive predictions for neutrino oscillations. 
Below we proceed with two different symmetry-breaking steps.

(i) A small perturbation to the (3,3) elements of $M_{0l}$
and $M_{0\nu}$ is introduced \cite{FH}. The resultant mass matrices
read
\begin{eqnarray}
M_{1l} & = & \frac{c^{~}_l}{3} \left ( \matrix{
1	& 1	& 1 \cr
1	& 1	& 1 \cr
1	& 1	& 1 + \varepsilon^{~}_l \cr } \right ) \; , \nonumber \\
M_{1\nu} & = & c_\nu \left ( \matrix{
1	& 0	& 0 \cr
0	& 1	& 0 \cr
0	& 0	& 1 + \varepsilon_\nu \cr } \right ) \; , 
\end{eqnarray}
where $|\varepsilon^{~}_l| \ll 1$ and $|\varepsilon_\nu| \ll 1$.
Now the charged lepton mass matrix ceases to be of rank one,
and the muon becomes massive ($m_\mu = 2|\varepsilon^{~}_l|
m_\tau /9$ to the leading order of $\varepsilon^{~}_l$). 
The neutrino mass $m_3$ is no more degenerate
with $m_1$ and $m_2$ (i.e., $|m_3 - m_0| = m_0 |\varepsilon_\nu|$). 
It is easy to see, after the diagonalization of 
$M_{1l}$ and $M_{1\nu}$, that the second and third
lepton families have a definite flavor mixing angle
$\theta$. We obtain $\tan\theta = -\sqrt{2} ~$ if the
small correction of $O(m_\mu/m_\tau)$ is neglected.
Then neutrino oscillations at the atmospheric scale arise
in $\nu_\mu \leftrightarrow \nu_\tau$ transitions with the
mass-squared difference $\Delta m^2_{32} = \Delta m^2_{31}
\approx 2m_0 |\varepsilon_\nu|$, where
$\Delta m^2_{ij} \equiv |m^2_i - m^2_j|$. The corresponding
mixing factor is in good agreement with current data
($\sin^2 2\theta \approx 8/9$).

(ii) A small perturbation to the (2,2) or (1,1) elements of
$M_{1l}$ and $M_{1\nu}$ is introduced, in order to
generate the electron mass and to lift the degeneracy between
$m_1$ and $m_2$. It has been argued in Refs. \cite{FX96,FTY98},
in analogy to the quark case,
that at this step a simple and instructive
perturbation to $M_{1l}$ should be of the form that
its (1,1) and (2,2) elements simultaneously receive
small corrections of the same magnitude and of the opposite
sign.
The analogous correction can be introduced to $M_{1\nu}$.
Then the mass matrices become
\begin{eqnarray}
M_{2l} & = & \frac{c^{~}_l}{3} \left ( \matrix{
1 -\delta_l	& 1	& 1 \cr
1	& 1 + \delta_l	& 1 \cr
1	& 1	& 1 + \varepsilon^{~}_l \cr } \right ) \; , \nonumber \\
M_{2\nu} & = & c_\nu \left ( \matrix{
1-\delta_\nu	& 0	& 0 \cr
0	& 1 + \delta_\nu	& 0 \cr
0	& 0	& 1 + \varepsilon_\nu \cr } \right ) \; , 
\end{eqnarray}
where $|\delta_l| \ll 1$ and $|\delta_\nu| \ll 1$.
One finds $m_e = |\delta_l|^2 m^2_\tau /(27 m_\mu)$ to the
leading order as well
as $m_1 = m_0 (1-\delta_\nu)$ and
$m_2 = m_0 (1+\delta_\nu)$. The diagonalization of 
$M_{2l}$ and $M_{2\nu}$ leads to a full $3\times 3$
flavor mixing matrix, given as $U_0$ in Eq. (2) if
small corrections of $O(\sqrt{m_e/m_\mu})$ and $O(m_\mu/m_\tau)$
are neglected.
Then the solar neutrino deficit can be interpreted by
$\nu_e \leftrightarrow \nu_\mu$ oscillations with
the mass-squared difference
$\Delta m^2_{21} \approx 4m_0 |\delta_\nu|$ 
and the maximal oscillation amplitude \cite{FX96,Sogami}.

If the corrections from nonvanishing muon and
electron masses are taken into account, the leptonic flavor
mixing matrix will in general read as $V = O_l U_0$, where
$O_l$ is an orthogonal matrix. Three rotation angles of $O_l$
are functions of the mass ratios
$m_e/m_\mu$ and $m_\mu/m_\tau$. Due to the strong hierarchy
of the charged lepton mass spectrum \cite{PDG98}, i.e.,
\begin{eqnarray}
\alpha & \equiv & \sqrt{\frac{m_e}{m_\mu}} \;
\approx 0.0695 \; , \nonumber \\
\beta & \equiv & \frac{m_\mu}{m_\tau} \; \approx \;
0.0594 \; ,
\end{eqnarray}
$O_l$ is expected not to deviate much from the unity matrix.
In our specific symmetry-breaking case discussed above,
we obtain 
\begin{equation}
O_l \; =\; \left ( \matrix{
1- \frac{1}{2} \alpha^2	& \alpha	& \sqrt{2} ~ \alpha \beta \cr
-\alpha	& 1-\frac{1}{2}\alpha^2 -\frac{1}{4}\beta^2
& -\frac{1}{\sqrt{2}} \beta \cr
-\frac{3}{\sqrt{2}} \alpha \beta	& \frac{1}{\sqrt{2}} \beta
& 1- \frac{1}{4} \beta^2 \cr } \right ) \;
\end{equation}
to the next-to-leading order. Note that there is
another 
solution for $O_l$ and it can directly be obtained from Eq. (7) with the
replacements $\alpha \rightarrow -\alpha$ and $\beta \rightarrow
-\beta$. 
The leptonic flavor mixing matrix turns out to be
\begin{equation}
V_{(\pm)} \; =\; U_0 \pm \left ( \alpha A -
\beta B \right ) - \left (\alpha^2 C - 
\alpha \beta D + \beta^2 E \right )\; ,
\end{equation}
in which the constant matrices $A \ldots E$ read as 
\begin{eqnarray}
A & = & \left ( \matrix{
\frac{1}{\sqrt{6}}	& \frac{1}{\sqrt{6}}	& -\frac{2}{\sqrt{6}} \cr
-\frac{1}{\sqrt{2}}	& \frac{1}{\sqrt{2}}	& 0 \cr
0	& 0	& 0 \cr } \right ) \; , 
\nonumber \\ \nonumber \\
B & = & \left ( \matrix{
0	& 0	& 0 \cr
\frac{1}{\sqrt{6}}	& \frac{1}{\sqrt{6}} 	& \frac{1}{\sqrt{6}} \cr
-\frac{1}{2\sqrt{3}}	& -\frac{1}{2\sqrt{3}} 
& \frac{1}{\sqrt{3}} \cr } \right ) \; ,
\nonumber \\ \nonumber \\ 
C & = & \left ( \matrix{
\frac{1}{2\sqrt{2}}	& -\frac{1}{2\sqrt{2}}	& 0 \cr
\frac{1}{2\sqrt{6}}	& \frac{1}{2\sqrt{6}}	& -\frac{1}{\sqrt{6}} \cr
0	& 0	& 0 \cr } \right ) \; , 
\nonumber \\ \nonumber \\
D & = & \left ( \matrix{
\frac{2}{\sqrt{6}}	& \frac{2}{\sqrt{6}}	& \frac{2}{\sqrt{6}} \cr
0	& 0	& 0 \cr
-\frac{3}{2}	& \frac{3}{2}	& 0 \cr } \right ) \; , 
\nonumber \\ \nonumber \\
E & = & \left ( \matrix{
0	& 0	& 0 \cr
\frac{1}{4\sqrt{6}}	& \frac{1}{4\sqrt{6}} 	& -\frac{1}{2\sqrt{6}} \cr
\frac{1}{4\sqrt{3}}	& \frac{1}{4\sqrt{3}} 
& \frac{1}{4\sqrt{3}} \cr } \right ) \; .
\end{eqnarray}
The effects of 
$O(\alpha^2)$, $O(\alpha \beta)$ and $O(\beta^2)$
on neutrino oscillations will be discussed subsequently.

In general a $3\times 3$ flavor mixing matrix $V$ can be parametrized,
in terms of three Euler angles and one $CP$-violating
phase, as follows \cite{FX97}:
\begin{equation}
V \; =\; \left ( \matrix{
s^{~}_l s_\nu c + c^{~}_l c_\nu	e^{-{\rm i}\phi}
& s^{~}_l c_\nu c - c^{~}_l s_\nu e^{-{\rm i}\phi}
& s^{~}_l s \cr
c^{~}_l s_\nu c - s^{~}_l c_\nu e^{-{\rm i}\phi}	
& c^{~}_l c_\nu c + s^{~}_l s_\nu e^{-{\rm i}\phi}
& c^{~}_l s \cr
-s_\nu s	& -c_\nu s	& c \cr } \right ) \; ,
\end{equation}
where $s^{~}_l \equiv \sin \theta_l$, $c_\nu \equiv \cos\theta_\nu$,
$s \equiv \sin\theta$, etc. Possible tiny $CP$-violating effects
will not be discussed here, and we take $\phi =0$.
We then obtain
$\tan\theta_l =0$, $\tan\theta_\nu =1$ and $\tan\theta =-\sqrt{2} ~$
in the limit where terms of 
$O(\alpha)$ and $O(\beta)$ are neglected. 
Taking small corrections of $O(\alpha)$ and $O(\beta)$ into account,
we arrive at $\tan\theta_l = \pm \alpha$, $\tan\theta_\nu = 1$ and
$\tan\theta = -\sqrt{2} ~ (1 \pm 3\beta/2)$, where the ``$\pm$''
signs correspond to $V_{(\pm)}$ in Eq. (8). 
The full next-to-leading-order
results for three mixing angles are found to be
\begin{eqnarray}
\tan\theta_l & = & \pm ~ \alpha
\left (1 \mp \frac{3}{2} \beta \right ) 
\;\; , \nonumber \\
\tan\theta_\nu & = & 1 ~ - ~ 3\sqrt{3} ~ \alpha \beta
\; , \nonumber \\
\tan\theta_{~} & = & -\sqrt{2} ~ \left ( 1 \pm \frac{3}{2}
\beta \right ) \; .
\end{eqnarray}
One can see that
the rotation angle $\theta_\nu$, which primarily describes  the mixing
between the first and second neutrino families,
only receives a tiny 
correction from the charged lepton sector.

Following Ref. \cite{FX96} we take 
$\Delta m^2_{\rm sun} = \Delta m^2_{21}$ and
$\Delta m^2_{\rm atm} = \Delta m^2_{32} \approx 
\Delta m^2_{31}$
to accommodate current data on solar and atmospheric neutrino oscillations.
Calculating the survival probability $P(\nu_e \rightarrow
\nu_e)$ and the transition probability $P(\nu_\mu \rightarrow
\nu_\tau)$ to the next-to-leading order, we arrive at
\begin{eqnarray}
\sin^2 2\theta_{\rm sun} & = & 1 ~ - ~ \frac{8}{3} \alpha^2
\; , \nonumber \\
\sin^2 2\theta_{\rm atm} & = & \frac{8}{9} \left (1 \mp
\beta \right ) \; .
\end{eqnarray}
The numerical results for mixing angles in Eqs. (11)
and (12) are listed in Table 1.
One can see that the flavor mixing patterns
``$V_{(+)}$'' and ``$V_{(-)}$'' are both consistent with
the present experiments.
\begin{table}[t]
\caption{Numerical results for mixing angles $\theta_l$,
$\theta_\nu$, $\theta$ and 
$\sin^2 2\theta_{\rm sun}$, $\sin^2 2\theta_{\rm atm}$
to the next-to-leading order.}
\vspace{-0.2cm}
\begin{center}
\begin{tabular}{cccccccc} \\ \hline\hline 
Case     &~~& $\theta_l$	& $\theta_\nu$	& $\theta$	
&~& $\sin^2 2\theta_{\rm sun}$ 	& $\sin^2 2 \theta_{\rm atm}$ 
\\  \hline \\
``$V_{(+)}$''	
&& $+3.6^{\circ}$	& $44.4^{\circ}$	& $-57.0^{\circ}$
&& $0.99$	& $0.84$ \\ \\
``$V_{(-)}$''
&& $-4.3^{\circ}$	& $44.4^{\circ}$	& $-52.2^{\circ}$
&& $0.99$	& $0.94$ \\  \\ 
\hline\hline
\end{tabular}
\end{center}
\end{table}

The near degeneracy of three neutrino masses assumed in the 
phenomenological scenario under discussion leads to
\begin{equation}
\left | \frac{m_2 - m_1}{m_3 - m_2} \right |
\; \approx \; \frac{\Delta m^2_{\rm sun}}{\Delta m^2_{\rm atm}}
\; \sim \; 10^{-7} \; ,
\end{equation}
or $|\delta_\nu|/|\varepsilon_\nu| 
\approx \Delta m^2_{\rm sun}/(2\Delta m^2_{\rm atm})\sim 10^{-7}$.
This kind of neutrino mass spectrum can account for the hot
dark matter of the universe, if $m_i \approx 2$ eV (for
$i=1,2,3$). The relatively large gap between $\Delta m^2_{21}$ and
$\Delta m^2_{32}$ (or $\Delta m^2_{31}$) has some implications
on the forthcoming long-baseline experiments, as we shall see
later on.

Now we consider the effect of nonvanishing $\theta_l$ on the
survival probability of electron neutrinos in a long-baseline
(LB) experiment, in which the oscillation associated with the
mass-squared difference $\Delta m^2_{21}$ can be safely 
neglected due to $\Delta m^2_{12} \ll \Delta m^2_{32}
\approx \Delta m^2_{31}$. It is easy to find
\begin{equation}
P (\nu_e \rightarrow \nu_e )^{~}_{\rm LB} \; =\;
1 ~ - ~ \frac{8}{3} \alpha^2
\left (1 \mp 2 \beta \right )
\sin^2 
\left (1.27 ~ \frac{\Delta m^2_{32} L}{|{\bf P}|} \right ) \; ,
\end{equation}
where ${\bf P}$ denotes the momentum of the neutrino beam
(in unit of GeV), and $L$ is the distance between the neutrino 
production and detection points (in unit of km).
The oscillation amplitude amounts to $1.1\%$ (the ``$V_{(+)}$'' case) 
or $1.4\%$ (the ``$V_{(-)}$'' case) and might be detectable.
Note that the CHOOZ experiment, 
in which the survival probability of anti-electron
neutrinos is measured, indicates the oscillation amplitude
$\sin^2 2\theta_{\rm CH} < 0.18$ for $\Delta m^2_{\rm CH}
\geq 9 \times 10^{-4} ~ {\rm eV}^2$ \cite{CHOOZ}. 
In the three-flavor scheme under
consideration, it is appropriate to set $\Delta m^2_{\rm CH} =
\Delta m^2_{32} \approx \Delta m^2_{31}$, which essentially
has no conflict with the Super-Kamiokande data \cite{BG98}.
One can see that the
small mixing obtained in Eq. (14) lies well within the allowed
region of $\sin^2 2 \theta_{\rm CH}$. 

The transition probability of $\nu_\mu$ to $\nu_e$ in such a
long-baseline neutrino experiment reads
\begin{equation}
P(\nu_\mu \rightarrow \nu_e)^{~}_{\rm LB} \; =\; 
\frac{16}{9} \alpha^2 \left (1 \mp
\beta \right ) \sin^2 
\left ( 1.27 ~ \frac{\Delta m^2_{32} L}{|{\bf P}|} \right ) \; .
\end{equation}
Here the mixing factor is about $0.8\%$ (the ``$V_{(+)}$'' case)
or $0.9\%$ (the ``$V_{(-)}$'' case).
The proposed K2K experiment is expected to have a sensitivity
of $\sin^2 2\theta > 10\%$ for $\nu_e \leftrightarrow
\nu_\mu$ oscillations, while the MINOS
experiment could probe values of the mixing as low as
$\sin^2 2\theta = 1\%$ \cite{MINOS}. Thus
a test of or a constraint on the prediction obtained in Eq. (15)
would be available in such experiments. 
On the other hand, the probability for
$\nu_\mu \rightarrow \nu_\tau$ transitions 
in the assumed long-baseline experiment is
essentially the same as that for the atmospheric neutrino oscillations,
i.e.,
\begin{equation}
P (\nu_\mu \rightarrow \nu_\tau)_{\rm LB} \; =\; 
\frac{8}{9} \left (1 \mp \beta \right )
\sin^2 \left ( 1.27 ~ \frac{\Delta m^2_{32} L}{|{\bf P}|} \right ) \; .
\end{equation}
The mixing factor,
corresponding to two different perturbative corrections of the magnitude
$\beta \sim 6\%$, takes the value 0.84 or 0.94 (see Table 1). 
It is also worth mentioning that the transition probability of 
$\nu_e \rightarrow \nu_\tau$, which satisfies
the sum rule
\begin{equation}
P(\nu_e \rightarrow \nu_e)_{\rm LB} \; +\;
P(\nu_e \rightarrow \nu_\mu)_{\rm LB} \; +\;
P(\nu_e \rightarrow \nu_\tau)_{\rm LB} \; =\; 1 \; ,
\end{equation}
is smaller (with the mixing factor $8\alpha^2/9 \approx
0.4\%$) and more difficult to detect.

As pointed out in Ref. \cite{FX96}, the three-flavor
scenario with
near degenerate neutrino masses and near-maximal
mixing angles has no conflict with current data on
the neutrinoless $\beta\beta$-decay \cite{Beta}, if neutrinos are
of the Majorana type. However, it is not compatible 
with the result of the LSND experiment \cite{LSND}.
The new analysis from the KARMEN experiment \cite{KARMEN}
seems to be in contradition with the LSND evidence for
$\bar{\nu}_\mu \leftrightarrow \bar{\nu}_e$ oscillations,
and a further examination of the latter will be available
in the coming years.

Finally let us give some comments on
the bi-maximal mixing scenario of three neutrinos,
which is recently proposed by Barger {\it et al} \cite{Barger}.
The relevant flavor mixing matrix, similar to 
$U_0$ in Eq. (2), reads as follows:
\begin{equation}
U' \; =\; \left (\matrix{
\frac{1}{\sqrt{2}}	& -\frac{1}{\sqrt{2}}	& 0 \cr
\frac{1}{2}	& \frac{1}{2}	& -\frac{1}{\sqrt{2}} \cr
\frac{1}{2}	& \frac{1}{2}	& \frac{1}{\sqrt{2}} \cr }
\right ) \; . 
\end{equation}
Such a flavor mixing pattern is independent of any lepton
mass and leads exactly to
$\sin^2 2\theta_{\rm atm} = \sin^2 2 \theta_{\rm sun} =1$
for neutrino oscillations. We find 
that $U'$ can be derived from the following charged lepton
and neutrino mass matrices:
\begin{eqnarray}
M'_l & = & \frac{c'_l}{2} \left [ 
\left ( \matrix{
0	& 0	& 0 \cr
0	& 1	& 1 \cr
0	& 1	& 1 \cr } \right ) + \left ( \matrix{
\delta'_l	& 0	& 0 \cr
0	& 0	& \varepsilon'_l \cr
0	& \varepsilon'_l	& 0 \cr } \right ) 
\right ] \; , \nonumber \\
M'_\nu & = & c'_\nu \left [ 
\left ( \matrix{
1	& 0	& 0 \cr
0	& 1	& 0 \cr
0	& 0	& 1 \cr } \right ) + \left ( \matrix{
0	& \varepsilon'_\nu 	& 0 \cr
\varepsilon'_\nu	& 0 	& 0 \cr
0	& 0	& \delta'_\nu \cr } \right ) \right ] \; , 
\end{eqnarray}
where $|\delta'_{l,\nu}| \ll 1$ and $|\varepsilon'_{l,\nu}|
\ll 1$. In comparison with the democratic mass matrix
$M_{0l}$ given in Eq. (3), which is invariant
under the $S(3)_{\rm L} \times S(3)_{\rm R}$ transformation, the
matrix $M'_l$ in the limit $\delta'_l = \varepsilon'_l =0$
only has the $S(2)_{\rm L} \times
S(2)_{\rm R}$ symmetry. However $M'_\nu$ 
in the limit $\delta'_\nu = \varepsilon'_\nu
=0$ takes the same form as $M_{0\nu}$, which displays the $S(3)$ symmetry.
The off-diagonal perturbation of
$M'_l$ allows the masses of three charged leptons to be
hierarchical:
\begin{equation}
\left \{ m_e \; , \; m_\mu \; , \; m_\tau \right \}
\; =\; \frac{c'_l}{2} \left \{ |\delta'_l| \; , \;
|\varepsilon'_l| \; , \; 2 + \varepsilon'_l \right \} \; . 
\end{equation}
We get $c'_l = m_\mu + m_\tau \approx 1.88$ GeV,
$|\varepsilon'_l| = 2m_\mu/(m_\mu + m_\tau) \approx 
0.11$ and $|\delta'_l| = 2m_e/(m_\mu + m_\tau) 
\approx 5.4 \times 10^{-4}$. The off-diagonal perturbation
of $M'_\nu$ makes three neutrino masses non-degenerate:
\begin{equation}
\left \{ m_1 \; , \; m_2 \; , \; m_3 \right \} 
\; =\; c'_\nu \left \{ 1+\varepsilon'_\nu \; , \;
1 - \varepsilon'_\nu \; , \; 1 + \delta'_\nu \right \} \; .
\end{equation}
Taking $\Delta m^2_{\rm sun} = \Delta m^2_{21}$ and
$\Delta m^2_{\rm atm} = \Delta m^2_{32} \approx \Delta m^2_{31}$
for solar and atmospheric neutrino oscillations, respectively,
we then arrive at $|\varepsilon'_\nu| /|\delta'_\nu|
\approx \Delta m^2_{\rm sun}/(2 \Delta m^2_{\rm atm})
\sim 10^{-7}$, a result similar to that obtained
above. The diagonalization of $M'_l$ and $M'_\nu$
leads straightforwardly to the flavor mixing matrix $U'$.
In Ref. \cite{Barger} a different neutrino
mass matrix has {\it reversely} been derived from the
given $U'$ in a flavor
basis that the charged lepton mass matrix is diagonal.
The emergence of the bi-maximal flavor mixing pattern from
$M'_l$ and $M'_\nu$ in Eq. (19) is, in our point of view, 
similar to 
that of the near-maximal flavor mixing pattern from $M_{2l}$ and
$M_{2\nu}$ in Eq. (5). 

In summary, we have displayed a simple and phenomenologically
instructive symmetry-breaking pattern for the charged lepton
mass matrix with flavor democracy and the neutrino mass
matrix with mass degeneracy. Large (near-maximal) leptonic
flavor mixing angles, which are favored by recent Super-Kamiokande
data on atmospheric and solar neutrino oscillations, emerge
naturally from our scenario and have been evaluated to the
next-to-leading order. The oscillation amplitudes of
$\nu_e \leftrightarrow \nu_e$ and $\nu_e \leftrightarrow \nu_\mu$
transitions are
predicted to be about $1\%$ for the upcoming long-baseline
neutrino experiments. 
We expect that further results from
the Super-Kamiokande and other neutrino
experiments could finally clarify if the
solar neutrino deficit is attributed to the long-wavelength
vacuum oscillation and provide stringent tests of the
model discussed here.

\vspace{0.5cm}
This work was supported in part by the CHRX-CT94-0579 program.



\begin{thebibliography}{99}

\bibitem{SK1} Super-Kamiokande Collaboration, 
Y. Fukuda {\it et al.}, Report No. hep-ex/9807003;
T. Kajita, 
talk given at the International Conference {\it Neutrino '98}, Takayama,
Japan, June 1998.

\bibitem{SK2} Y. Suzuki, talk given at the International
Conference {\it Neutrino '98}, Takayama, Japan, June 1998;
Super-Kamiokande Collaboration, 
Y. Fukuda {\it et al.}, Report No.
hep-ex/9803006, hep-ex/9805006;
Y. Totsuka, talk given at the 18th International Symposium on
Lepton-Photon Interactions, Hamburg, Germany, July 1997.

\bibitem{FX96} H. Fritzsch and Z.Z. Xing, Phys. Lett. B {\bf 372}
(1996) 265;
Report No. LMU-98-07, hep-ph/9807234 (talk given at the Ringberg
Euroconference on New Trends in Neutrino Physics, Ringberg,
Germany, May 1998).

\bibitem{Smirnov} S.T. Petcov and A.Y. Smirnov,
Phys. Lett. B {\bf 322} (1994) 109;
Z.Z. Xing, Report No. DPNU-98-12, hep-ph/9804433 
(talk given at the Workshop on Fermion Masses
and $CP$ Violation, Hiroshima, Japan, March 1998).

\bibitem{FTY98} M. Fukugita, M. Tanimoto, and
T. Yanagida, Phys. Rev. D {\bf 57} (1998) 4429;
M. Tanimoto, Report No. hep-ph/9807577;
Report No. hep-ph/9807283.

\bibitem{FH} H. Fritzsch and D. Holtmannsp$\rm\ddot{o}$tter,
Phys. Lett. B {\bf 338} (1994) 290.

\bibitem{Sogami} See, also, I.S. Sogami, H. Tanaka, and
T. Shinohara, Report No. hep-ph/9807449.

\bibitem{PDG98} Particle Data Group,  
Eur. Phys. J. C {\bf 3} (1998) 1.

\bibitem{FX97} H. Fritzsch and Z.Z. Xing, Phys. Lett. B {\bf 413}
(1997) 396; Phys. Rev. D {\bf 57} (1998) 594.

\bibitem{CHOOZ} CHOOZ Collaboration, M. Apollonio {\it et al.},
Phys. Lett. B {\bf 420} (1998) 397.

\bibitem{BG98} S.M. Bilenky and C. Giunti, Report No.
hep-ph/9802201.

\bibitem{MINOS} Y. Totsuka, private communication;
P.Osland and G. Vigdel, Report No. hep-ph/9806339;
K. Zuber, Report No. hep-ph/9807468.

\bibitem{Beta} H.V. Klapdor-Kleingrothaus, Report No. hep-ex/9802007;
and references therein.

\bibitem{LSND} C. Athanassopoulos {\it et al.}, Phys. Rev.
Lett. {\bf 77} (1996) 3082.

\bibitem{KARMEN} B. Zeitnitz, talk given at the International
Conference {\it Neutrino '98}, Takayama, Japan, June 1998.

\bibitem{Barger} V. Barger, S. Pakvasa, T.J. Weiler, and 
K. Whisnant, Report No. hep-ph/9806387.
\end{thebibliography}
\end{document}